\documentclass[preprint,showpacs,showkeys,aps,prb]{revtex4}
\usepackage[english]{babel}

\usepackage[dvips]{graphicx}
\usepackage{amssymb}

\begin{document}


\title{Well-Posed Two-Temperature Constitutive Equations for Stable
Dense Fluid Shockwaves using Molecular Dynamics and
Generalizations of Navier-Stokes-Fourier Continuum Mechanics
}

\author{Wm. G. Hoover and Carol G. Hoover \\
Ruby Valley Research Institute \\ Highway Contract 60,
Box 598, Ruby Valley 89833, NV USA}

\date{\today}



\vskip 0.5cm

\begin{abstract}

Guided by molecular dynamics simulations, we generalize the
Navier-Stokes-Fourier constitutive equations and the continuum motion
equations to include both transverse and longitudinal temperatures. 
To do so we partition the contributions of the heat transfer, the work
done, and the heat flux vector between the longitudinal and transverse
temperatures.
With shockwave boundary conditions time-dependent solutions of these
equations converge to give stationary shockwave profiles. The profiles
include anisotropic temperature and can be fitted to molecular dynamics
results, demonstrating the utility and simplicity of a two-temperature
description of far-from-equilibrium states.


\end{abstract}

\maketitle

\begin{figure}
\includegraphics[height=8cm,width=2cm,angle=-90]{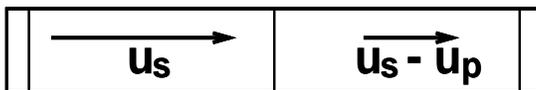}
\caption{
Schematic stationary shockwave.  Cold fluid enters at the left cold boundary,
with speed $u_s$; hot fluid leaves at the right hot boundary, with speed
$u_s-u_p$.  We choose a coordinate frame which moves leftward, at speed $u_s$
relative to the laboratory frame.  The shockwave remains stationary in this
coordinate frame.
}
\end{figure}

\begin{figure}
\includegraphics[height=6cm,width=4.5cm,angle=-90]{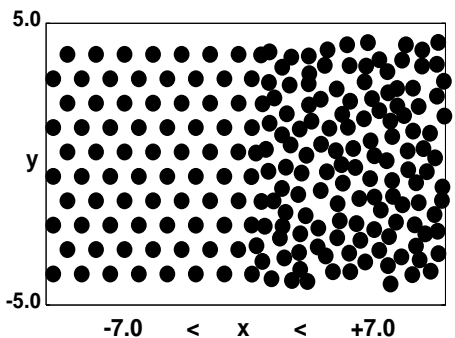}
\caption{
Stationary shockwave.  Snapshot from a 10-row molecular dynamics
simulation with a periodic height of $10\sqrt{3/4}$.  The simulations
analyzed in the text are based on 80-row molecular dynamics with a
periodic height of $80\sqrt{3/4}$.
}
\end{figure}

\section{Introduction}

A leftmoving piston, impacting a fluid with velocity $-u_p$, generates a
leftmoving shockwave with velocity $-u_s$.  Throughout this paper we
analyze such a shockwave from the viewpoint of a coordinate system moving
leftward, so as to keep pace with the shock.  See Figures 1 and 2.  In this
special uniformly-translating coordinate frame the shockwave is stationary,
simplifying theoretical analyses.  One-dimensional stationary
shockwaves\cite{b1,b2,b3,b4,b5,b6,b7,b8,b9,b10,b11,b12,b13,b14}
provide a useful computational laboratory for the study of stationary
far-from-equilibrium states.  In such a shockwave a cold fluid is converted
irreversibly to a hot one.  As the fluid moves from left to right, in the
shock-centered coordinate frame of the Figures,
at speed $u(x)$, the $x$ coordinate increases; typically, the corresponding
density, the longitudinal component of the pressure tensor, and the energy
all increase too, in just such a way that the spatial structure of the wave
is stationary:
$$
\{ \ u = \dot x,\dot \rho,\dot P_{xx},\dot e  \ \} > 0 \ ,
$$
$$
(\partial u/\partial t)_x = 0 \ ; \
(\partial \rho/\partial t)_x = 0 \ ; \
(\partial P_{xx}/\partial t)_x = 0 \ ; \
(\partial P_{yy}/\partial t)_x = 0 \ ; \
(\partial e/\partial t)_x = 0 \ .
$$
As the velocity decreases from its leftmost entrance value,
$u(x \rightarrow -\infty) = u_s$,
to its rightmost exit value,
$u(x \rightarrow +\infty) = u_s-u_p$,
the stationary nature of the wave requires that the fluxes of mass, momentum,
and energy remain constant throughout:
$$
(\rho u)_x = (\rho u)_{\rm cold} = (\rho u)_{\rm hot} \ ;
$$
$$
(P_{xx} + \rho u^2)_x = (P+\rho u^2)_{\rm cold} = (P+\rho u^2)_{\rm hot} \ ;
$$
$$
(\rho u)[(e+(P_{xx}/\rho)+(u^2/2)]_x +Q_x =
$$
$$
(\rho u)[e+(P/\rho)+(u^2/2)]_{\rm cold} = 
(\rho u)[e+(P/\rho)+(u^2/2)]_{\rm hot} \ .
$$
The notation here is conventional, with the pressure tensor $P$ and heat
flux vector $Q$ assumed to be calculable from the density $\rho$,
velocity $u$, energy $e$, and their gradients.

Temperature\cite{b11,b12,b15,b16,b17} is our special interest in this work.
Temperature is most simply and usefully defined as a velocity fluctuation,
the ``kinetic temperature'':
$$
kT_{xx} \equiv m\langle (\dot x - \langle \dot x \rangle)^2 \rangle \ ; \
kT_{yy} \equiv m\langle (\dot y - \langle \dot y \rangle)^2 \rangle \ .
$$
The angular brackets imply a local average.  The velocities here are
individual particle velocities, whose local average would be the hydrodynamic
flow velocity $u$.  Temperature is just the fluctuation about this average.
It is evident that $T_{xx}$ and $T_{yy}$ can differ. In dilute-gas
kinetic theory, the difference corresponds to a shear stress:
$$
\rho k(T_{xx} - T_{yy})/(2m) = 
(P_{xx} - P_{yy})/2 \ \  {\rm [Dilute \ Gas]} \ ,
$$
where $k$ is Boltzmann's constant and $m$ is the particle mass, which
we choose equal to unity in what follows. In dense fluids there is no simple
relationship between the two tensors so that special evolution equations
for $T_{xx}$ and $T_{yy}$ need to be developed, as we do in Section III.

The cold fluid, initially
moving to the right at the entrance velocity, or ``shock velocity'' $u_s$,
is slowed by its encounter with the wave until it reaches its exit
velocity $u_s - u_p$, where $u_p$ is the ``piston velocity'' or
``particle velocity''.  In this irreversible deceleration the kinetic
energy lost by the decelerating fluid is converted into additional hot
fluid enthalpy $(H = E + PV \leftrightarrow h = e + Pv)$:
$$
h_{\rm hot} - h_{\rm cold} =
[e +(P/\rho)]_{\rm hot}-[e +(P/\rho)]_{\rm cold} =
[\rho_{\rm cold} u_s^2/2] - [\rho_{\rm hot} (u_s-u_p)^2/2] \ .
$$ 
The cold and hot boundary conditions enclosing the shock are equilibrium ones 
imposed far from the shockfront so that the small-system surface effects
complicating the number-dependence of nonequilibrium systems are minimized.
In implementing these ideas no arbitrary or artificial assumptions
have to be made.  All the observed phenomena follow from the assumed form
for the interparticle forces.  Figures 3, 4, and 5 show typical results from
molecular dynamics, as is described in more detail in Section II.  Notice
that the rise in longitudinal temperature $T_{xx}$ can be much larger
and can occur somewhat earlier\cite{b12} than that of the transverse
temperature $T_{yy}$.

\begin{figure}
\includegraphics[height=9cm,width=5.25cm,angle=-90]{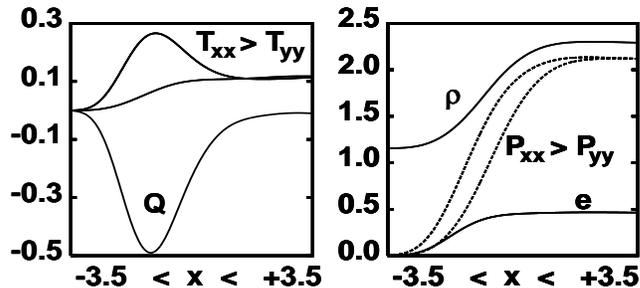}
\caption{
A snapshot spatial profile of a nominally steady one-dimensional shockwave
from molecular dynamics, using a short-ranged repulsive potential.
Spatial one-dimensional averages of the temperatures and heat flux (left)
and the pressures, density, and energy (right) have been computed with
Lucy's weight function using a range $h=3$.  The cold
zero-pressure, zero-temperature triangular lattice is compressed to twice the
initial density ($\sqrt{4/3} \rightarrow 2\sqrt{4/3}$) by the shockwave, just
as in Figure 2.
}
\end{figure}

\begin{figure}
\includegraphics[height=9cm,width=5.25cm,angle=-90]{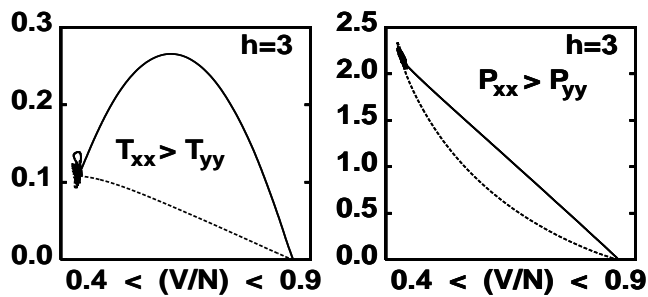}
\caption{
Volume dependence of the temperature tensor (left) and the pressure tensor
(right) in the stationary shockwave of Figure 3, as calculated with
molecular dynamics.  Spatial averages have been computed with Lucy's weight
function using a range $h=3$, as is discussed in Section II.
}
\end{figure}

\begin{figure}
\includegraphics[height=9cm,width=5.25cm,angle=-90]{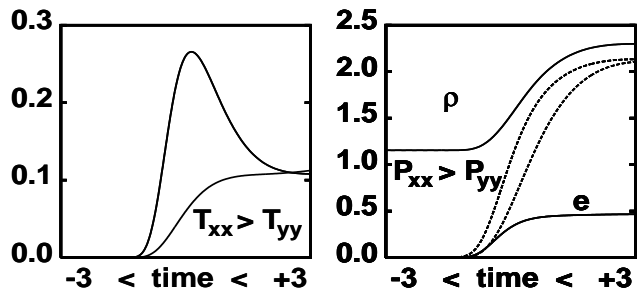}
\caption{
Stationary temporal profile for the one-dimensional shockwave of
Figure 3, using a short-ranged repulsive potential.
Spatial averages of the temperatures (left) and the pressures, density,
and energy (right) have been computed with Lucy's weight function using a
range $h=3$.  The initial
stress-free cold triangular lattice is compressed to twice the initial
density by the shockwave, as in Figure 2.  The time origin has been chosen,
arbitrarily, close to the shockfront.
}
\end{figure}

In Section III we discuss the {\em continuum} mechanics of the same
shockwave problem.  Evidently {\em any} continuum formulation must first
of all include the continuum conservation laws for mass, momentum, and
energy:
$$
\dot\rho = -\rho \nabla \cdot u \ ;
$$
$$
\rho \dot u = -\nabla \cdot P \ ;
$$
$$
\rho \dot e = -\nabla u:P - \nabla \cdot Q \ .
$$
Here the pressure tensor $P$ and heat flux vector $Q$ measure the momentum
and energy fluxes in the local ``comoving'' (or ``Lagrangian'') coordinate
frame moving with the mean velocity $u(x)$.  Now the superior dot notation
is used to indicate the time derivatives of $\rho$, $u$, and $e$ following
the motion at velocity $u$.  In the continuum description these field
variables are continuous differentiable functions of space and time so that
the spatial averaging (necessary to an analysis of molecular dynamics data)
is unnecessary.

The steady nature of the shock process makes it possible to use either
space or time as an independent variable.  On the average, the progress
of a particle traveling through the shockwave follows from the integral
of the flow velocity. To illustrate, consider
again the molecular dynamics profiles shown in Figures 3, with space
as the abscissa.  Exactly the same profiles can
alternatively be expressed with time as the abscissa, as in Figure 5.
To change from space-based to time-based profiles requires use of the
ratio $(dx/dt) \equiv u$:
$$
\int_0^t dt^\prime = \int_{x_0}^x dx^\prime/u(x^\prime) \ ; \
t = 0 \leftrightarrow x = x_0 \ .
$$
where $u(x)$ is the hydrodynamic flow velocity.  Thus all the spatial
snapshots or equivalent temporal wave profiles
catalog the sequence of time-ordered states through which the particles
in a typical volume (initially at $x_0$) pass as they transit the shockwave.

Because the conventional Navier-Stokes-Fourier approach, illustrated
in Figure 6, assumes a scalar temperature, $T = T_{xx} = T_{yy}$, several
modifications of the continuum description need to be made to model the
two-temperature results of Figures 3-5 found with molecular dynamics, with
$T_{xx} \neq T_{yy}$.  In Section III we describe simple modifications of the
Navier-Stokes-Fourier constitutive and flow equations, along with a numerical
method which converges nicely to give stationary shockwave profiles in the
two-temperature case.

\begin{figure}
\includegraphics[height=9cm,width=5.25cm,angle=-90]{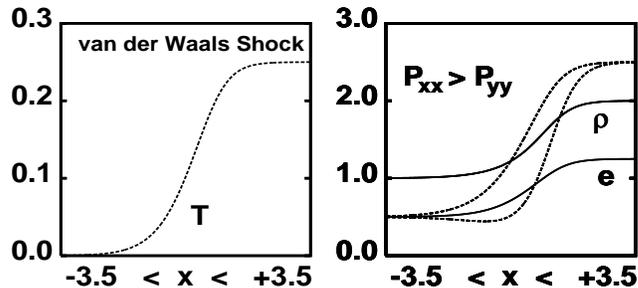}
\caption{
Stationary spatial profile for a one-dimensional shockwave
according to the usual Navier-Stokes-Fourier equations for
the model fluid: $P_{\rm eq} = \rho e \ ; \ e = (\rho /2) + kT$ with
unit shear viscosity, zero bulk viscosity, and unit Fourier heat
conductivity.  Here the temperature $T$ (left) is a scalar, as in conventional
continuum mechanics.
}
\end{figure}

Section IV is reserved for a summary and our concluding remarks, including
suggestionns for adapting our ideas to detailed two- and three-dimensional
descriptions of the fluctuations in nonequilibrium systems.

\section{Results from Molecular Dynamics}

The molecular dynamics simulations leading to our current results are all
based on a very simple model two-dimensional system of unit-mass 
unit-radius particles interacting pairwise with a short-ranged normalized
repulsive potential\cite{b12,b15}:
$$
\phi(r<1) = (10/\pi)(1 - r)^3 \ \rightarrow \ 
\int_0^12\pi rdr\phi(r) \equiv 1 \ .
$$
The length and energy scales set by this potential correspond to the range
and strength of the interparticle pair forces.
The equilibrium properties for this potential can be approximated very
roughly by a theoretical model (based on a random distribution of
particles in space) resembling van der Waals' mean-field idea,
$$
P = \rho e \ ; \ e = (\rho/2) + kT \ .
$$
$P$, $\rho$, $e$, and $T$ are the pressure, density, energy, and 
temperature. Though the models and language here all refer to
systems in two space dimensions the same ideas can be applied equally
well to three-dimensional systems.

We expect that the nonequilibrium properties for
this model will likewise provide a simple interpretation.  We are
particularly interested here in generalizing the notion of
temperature to the tensor case, $T_{xx} \neq T_{yy}$.  The need
for this generalization stems from the molecular dynamics shockwave
simulations summarized in Figures 3, 4, and 5.

Stationary shockwaves were obtained from molecular dynamics by matching
the mass flux of a cold stress-free lattice ($\rho = \sqrt{4/3}$ and
speed 1.930) to the mass flux of the hot fluid exiting at the righthand
boundary (with $\rho = 2\sqrt{4/3}$ and speed 0.965):
$$
\rho u = \rho _{\rm cold}u_{\rm cold} = \rho _{\rm hot}u_{\rm hot} =
1.93\times\sqrt{4/3} = 2.229 \ .
$$
With this choice for the shockwave speed $u_s = 1.93$ and particle (or
piston) speed $u_p = u_s/2$ the shockwave is stationary and corresponds
to twofold compression, a ``strong'' shockwave\cite{b12}. The Mach
number $M=u/c_s$ is not a useful description here as the sound speed
$c_s$ vanishes in the cold state.   The momentum and energy fluxes throughout
the wave are equal to those of the initial cold lattice:
$$
P_{xx} + \rho u^2 = \sqrt{4/3}(1.93)^2 = 4.301 \ ;
$$
$$
\rho u[e + (P_{xx}/\rho ) + (u^2/2)] + Q_x = 
\sqrt{4/3}(1.93)^3/2 = 4.151 \ .
$$

Spatial averages within the shockwave were calculated here using Lucy's
weight function\cite{b12,b13,b15,b16},
$$
w_{Lucy}(|x|<h) = (5/4h)[1 - 6r^2 + 8r^3 - 3r^4] \ ; \ r \equiv |x|/h < 1 \ ,
$$
with a range equal to three times the range of the potential, $h=3$.  The
internal energy at a gridpoint coordinate $x$, for example, is computed as
a ratio of sums:
$$
e(x) = \frac{\sum_i w(x - x_i)e_i}{ \sum_i w(x - x_i)} \ ,
$$
where the energy of Particle $i$ is the sum of its kinetic energy relative
to the local flow velocity $u(x)$ plus half its summed-up interaction
energy with other nearby Particles $\{j\}$.

Consider now the results shown in Figures 3 and 4.  The density, energy, and
pressure agree roughly with the hyperbolic-tangent profiles derived by Landau
and Lifshitz for a weak shockwave with constant transport coefficients\cite{b3}.
Figure 4 shows the pressure-temperature-volume states through which the
moving fluid travels.  The Rayleigh Line, a straightline relation linking
$P_{xx}$ and the volume, is necessarily satisfied and corresponds to the
conservation of momentum.  In marked contrast, the molecular dynamics
temperature shows a strong maximum (as might be expected from the mixing
of cold and hot Gaussian distributions suggested by Mott-Smith\cite{b1})
at the shockfront. Because the work done in compressing the fluid appears
first in the longitudinal direction we expect that the rise in $T_{xx}$
precedes that of $T_{yy}$, as is confirmed in Figure 3. This thermal
anisotropicity differs from the conventional textbook result and is the
main motivation for our work on a two-temperature continuum description,
detailed in the following Section.

\section{Results from Continuum Mechanics}

\subsection{General Considerations}

Continuum models combine the universal conservation laws (mass, momentum,
and energy) and the corresponding evolution equations (continuity, motion,
and energy) with specific constitutive models.  The constitutive models
describe the pressure tensor and the heat flux vector for nonequilibrium
systems.  The usual Navier-Stokes assumptions, which we follow here for a
two-dimensional fluid, are that the pressure tensor and heat flux vector
respond linearly to velocity and temperature gradients:
$$
P = P^{\rm eq} - \lambda[\nabla \cdot u]I - \eta [\nabla u + \nabla u^t] \ ; \
\lambda \equiv \eta _V - \eta \ .
$$
$$
Q = -\kappa \nabla T \ .
$$
It needs to be emphasized that the choice of particular expansion variables,
here $\nabla u$ and $\nabla T$, affects the solutions of nonlinear problems
like shockwave structure.  Garc\'ia-Col\'in and Green emphasized that the
description of nonequilibrium continuum mechanics is ambiguous whenever
the choice of ``equilibrium'' variables -- energy or longitudinal
temperature or transverse temperature in this case -- is ambiguous\cite{b17}. 
The numerical value of a Taylor's series in the deviations from equilibrium,
truncated after the first nonlinear term, is clearly sensitive to the
choice of independent variable.

In the nonequilibrium pressure tensor the superscript $^t$ indicates the
transposed tensor and $I$ is the unit tensor:
$$
I_{11} = I_{22} = 1 \ ; \ I_{12} = I_{21} = 0 \ ,
$$
$\eta $ is the shear viscosity, and $\lambda = \eta _v - \eta $ is defined
by the bulk viscosity $\eta _v$.  In the shockwave problem the pressure-tensor
definitions give
$$
P_{xx} = P^{\rm eq} - (\eta _v + \eta )du/dx \ ; \
P_{yy} = P^{\rm eq} - (\eta _v - \eta )du/dx \ .
$$
For a two-temperature continuum model it is necessary to formulate the
``equilibrium pressure'' $P^{\rm eq}$ as a function of the (nonequilibrium)
energy, density, and the two temperatures.  The viscosities and conductivity
could likewise depend upon these state variables and $\kappa$ can be a tensor,
as we show later, with an example.

When we define $T_{xx}$ and $T_{yy}$ as continuum state variables it becomes
necessary for us to formulate constitutive relations for their evolution.
The simplest such models begin by separating the energy into two parts: a
density-dependent ``cold curve'' $e^{\rm cold}(\rho)$ and an additional
kinetic or ``thermal'' part, proportional to temperature:
$$
e \equiv e^{\rm cold} (\rho ) + e^{\rm thermal}(T_{xx},T_{yy}) = 
e^{\rm cold} + (ck)(T_{xx} + T_{yy}) \ ,
$$
where $ck$ is a scalar heat capacity.  The functional form of the cold curve
produces a corresponding contribution to the pressure:
$$
P^{\rm cold} = -de^{\rm cold}/d(V/N) = \rho ^2 de^{\rm cold}/d\rho \ .
$$
Gr\"uneisen's $\gamma $ defines a corresponding thermal pressure:
$$
P^{\rm thermal} = \gamma \rho e^{\rm thermal}.
$$
The viscous part of the pressure tensor is Newtonian:
$$
P^{\rm viscous} = -\lambda \nabla \cdot uI - \eta (\nabla u + \nabla u^t) \ .
$$

The thermal and viscous parts of the First-Law energy change are then
apportioned between the $x$ and $y$ directions so as to be consistent
with overall energy conservation:
$$
\dot e^{\rm thermal} = \dot e - \dot e^{\rm cold} (\rho ) =
ck\dot T_{xx} + ck\dot T_{yy} \ ;
$$
$$
\rho ck\dot T_{xx} = -\alpha \nabla u:(P - IP^{\rm cold})
- \beta \nabla \cdot Q + \rho ck(T_{yy} - T_{xx})/\tau \ ; 
$$
$$
\rho ck\dot T_{yy} = (\alpha - 1) \nabla u:(P - IP^{\rm cold})
+ (\beta - 1) \nabla \cdot Q + \rho ck(T_{xx} - T_{yy})/\tau \ .
$$
The thermal relaxation time $\tau $ has been introduced in the evolution
equations to guarantee thermal equilibrium far from the shockwave:
$$
K_x = K_y \leftrightarrow T_{xx} = T_{yy} = T^{\rm eq} \ .
$$

In what follows we consider two models for the cold curve and the heat
capacity.  First, a weak repulsive pair force suggests implementing a
``van der Waals model'':
$$
e^{\rm cold} = (\rho /2) \ ; \
e^{\rm thermal} = k(T_{xx} + T_{yy})/2 \ ; \
P^{\rm eq}  = \rho e
$$
Second, a triangular-lattice-based model, based on Gr\"uneisen's ideas,
uses the nearest-neighbor static lattice energy and pressure corresponding
to the pair potential evaluated at the nearest-neighbor lattice spacing
$r$, $\phi = (10/\pi)(1-r)^3$:
$$
e^{\rm cold} = (30/\pi)(1-r)^3 \ ; \ 
p^{\rm cold}(V/N) = (45/\pi)r(1-r)^2 \ ;
$$
$$
r = \sqrt{V/V_0} \ ; \ V_0 = \sqrt{3/4}N \ .
$$
The corresponding equilibrium equation of state separates the energy and pressure
into ``cold'' and ``thermal'' parts:
$$
e^{\rm eq} = e^{\rm cold} + e^{\rm thermal} \ ; \
P^{\rm eq} = P^{\rm cold} + \rho \gamma e^{\rm thermal} \ ,
$$
with $\gamma $ chosen so as to roughly reproduce equation of state data
from molecular dynamics.  Let us next apply these two simple cold-curve models to
the shockwave problem.

\subsection{Potential plus Kinetic van der Waals Models}

First consider an arbitrary, but simple and natural, choice:
$$
P^{\rm eq} = \rho e \ ; \ e^{\rm eq} =  e^{\rm cold} + e^{\rm thermal} =
(\rho + kT_{xx} + kT_{yy})/2 \ .
$$
$$
P^{\rm cold} = \rho e^{\rm cold} = \rho ^2/2 \ ,
$$
with an initial density of unity and an initial temperature of zero.
Twofold compression of the cold van der Waals fluid gives the following
solution relating the initial and final equilibrium states:
$$
\rho : 1 \rightarrow 2 \ ; \
u    :     2 \rightarrow 1 \ ; \
T    :     0 \rightarrow 1/4 \ ; \
e    :     1/2 \rightarrow 5/4 \ ; \
P    :     1/2 \rightarrow 5/2 \ .
$$
The mass, momentum, and energy fluxes connecting these states must be
constant throughout the profile:
$$
\rho u = 2 \ ; \
P_{xx} + \rho u^2 = 9/2 \ ; \
\rho u[e + (P_{xx}/\rho) + (u^2/2)] + Q_x  = 6 \ .
$$

Consider the most extreme anisotropic situation consistent with energy
conservation, in which all the work done and heat transfered are
associated with thermal change in the $x$ direction.  The thermal
relaxation time $\tau $, here chosen equal to unity, guarantees that
the $x$ and $y$ temperatures equilibrate in a time of order $\tau $:
$$
\dot e^{\rm thermal} = \dot e - \dot e^{\rm cold} (\rho ) =
(k/2)(\dot T_{xx} + \dot T_{yy}) \ ;
$$
$$
\rho (k/2)\dot T_{xx} = -\nabla u:(P - IP^{\rm cold})
- \nabla \cdot Q + \rho (k/2)(T_{yy} - T_{xx})/\tau \ ; 
$$
$$
\rho (k/2)\dot T_{yy} = \rho (k/2)(T_{xx} - T_{yy})/\tau \ ; \ \tau = 1 \ .
$$
Two solutions of these equations appear in Figures 7 and 8.  For both
of them we chose a shear viscosity of unity and a vanishing bulk viscosity:
$$
P_{xx} = P^{\rm eq} - du/dx \ ; \ P_{yy} = P^{\rm eq} + du/dx \ .
$$

The heat flux vector requires that an additional choice be made for its
response to the gradients of $T_{xx}$ and $T_{yy}$.  We compare two
choices in Figures 7 and 8.  For both of them the overall conductivity
is unity, but the heat flux responds differently to the two components
of $\nabla T$:
$$
Q_x = -\kappa \nabla T_{yy} = \ -\nabla T_{yy} \ \ [{\rm Choice} \ 1] \ .
$$
$$
Q_x = -\kappa (\nabla T_{xx} + \nabla T_{yy})/2 =
-(\nabla T_{xx} + \nabla T_{yy})/2\ \ [{\rm Choice} \ 2] \ ;
$$

It is good fortune that the shockwave equations we summarize here
are relatively easy to
solve numerically.  The usual numerical method is the ``backward Euler''
scheme\cite{b2}.  One starts near the ``hot'' boundary and integrates
backward, using a first-order difference scheme.  That approach fails
here, due to the temperature relaxation terms, which are exponentially
unstable in the time-reversed case.  An integration forward in time is
required in the presence of relaxation.  A successful ``staggered-grid''
(two separate spatial grids) algorithm results if the density $\rho_c$
is defined at cell centers and energy, temperature, and pressure are
defined at the nodes which bound the cells\cite{b18,b19}.  This algorithm
follows the dynamics correctly and converges nicely to the stationary
profiles shown in Figures 7 and 8. A computational mesh spacing of $dx
= 0.1$ is sufficient, using the second-order spatial differencing scheme
outlined in References 18 and 19 with fourth-order Runge-Kutta time
integration.

In the early days of shockwave modeling this computational simplicity
was by no means apparent, so that there is an abundant literature on
the stability of numerical methods for the shockwave problem\cite{b2}.
Now, in the early days of tensor-temperature models, the main challenge
is to develop well-posed constitutive equations consistent with both
the conservation laws and the empirical results from molecular dynamics.

\begin{figure}
\includegraphics[height=9cm,width=5.25cm,angle=-90]{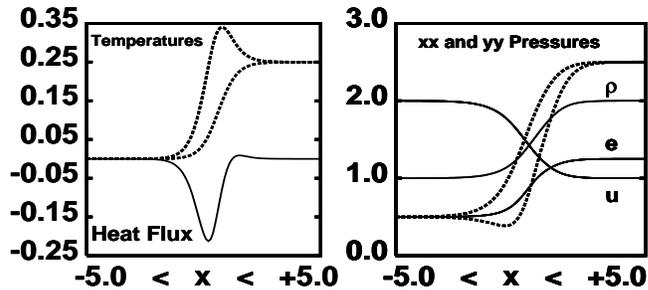}
\caption{
Typical solution of the {\em generalized} Navier-Stokes-Fourier equations
for the van der Waals model
with both heat and work contributing to $T_{xx}$ and with the heat flux
responding only to the gradient of $T_{yy}$.  The shear viscosity, heat
conductivity, heat capacity, and thermal relaxation times are all taken
equal to unity.
}
\end{figure}

\begin{figure}
\includegraphics[height=9cm,width=5.25cm,angle=-90]{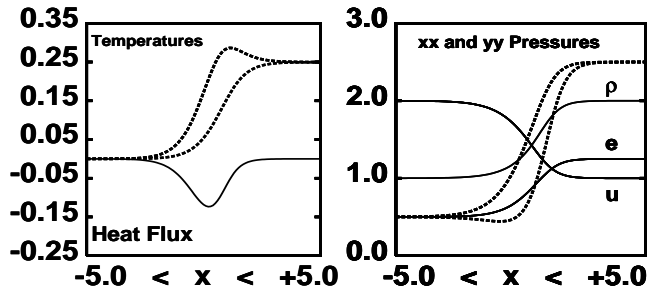}
\caption{
Typical solution of the {\em generalized} Navier-Stokes-Fourier equations
for the van der Waals model
with both heat and work contributing to $T_{xx}$ and with the heat flux
responding equally to the gradients of both $T_{xx}$ and $T_{yy}$. The
shear viscosity, heat conductivity, heat capacity, and thermal relaxation
times are all taken equal to unity.
}
\end{figure}

Interesting aspects of both solutions are (i) the minimum in $P_{yy}(x)$,
which suggests the need for bulk viscosity in modeling molecular dynamics
results, and (ii) the pronounced maximum in $T_{xx}(x)$, leading the response
of $T_{yy}$ and roughly equal in magnitude to that found in the dynamical
results of Section II.

The physical ideas incorporated in this simplest approach are four: (i) the
pressure and the work done can usefully be separated into a ``cold'' part and
a ``thermal'' part; (ii) the heat flux $Q$ responds to a linear combination
of the temperature gradients $\nabla T_{xx}$ and $\nabla T_{yy}$ in the usual
way, supplemented by (iii) the thermal relaxation of the thermal
anisotropicity, and (iv) separate linear combinations of the work done and heat
absorbed contribute to $T_{xx}$ and $T_{yy}$ throughout the shock compression
process.  

Here the total pressure, $P = P^\Phi + P^K$, contains potential and kinetic
components, measurable separately with molecular dynamics.  These extensions
of the Navier-Stokes approach closely parallel the relaxation-time treatments
of strong ideal-gas shockwaves carried out by Xu, Josyula, Holian, and
Mareschal\cite{b11,b14}.  Our more general approach necessarily differs from
theirs by allowing for contributions from the potential energy to temperature
changes and the transfer of heat.  The pressure profiles shown in Figures 7
and 8 also indicate the need for bulk viscosity, in that the molecular
dynamics results show a monotone-increasing $P_{yy}$, in contrast to the
distinct minimum found here in the absence of bulk viscosity.  We turn next to
a slightly more sophisticated model, an extension of Gr\"uneisen's equilibrium
equation of state.

\subsection{Cold plus Thermal Gr\"uneisen Models}

For gases, where the pressure and temperature tensors are proportional to
one another, a systematic expansion of the Boltzmann equation can be, and
has been, tried\cite{b10,b11,b14,b17}.  Xu and Josyula\cite{b11} as well
as  Holian and Mareschal\cite{b14} developed solutions of generalized
relaxation-time Boltzmann equations for the shockwave problem. For
dense fluids only Enskog's hard-sphere-based theory is available.  More
flexible empirical models need to be developed for dense-fluid shockwaves. 
A trial set of two-temperature evolution equations, the simplest plausible
set generalizing the van der Waals model above, makes use of Gr\"uneisen's
``cold curve'' representation of the energy and pressure to define ``thermal''
contributions.  These thermal parts include both the effects of thermal
agitation (heat and temperature) and  of mechanical distortion (work, through
compression with viscous deformation):
$$
E = \Phi^{\rm cold} + E^{\rm thermal} \ ; \
P_{(xx \ {\rm and} \ yy)} = P^{\rm cold} + P^{\rm thermal} + P^{viscous}\ .
$$

For the molecular dynamics simulations discussed in Section II
the cold parts of the pressure and energy, as well as their
time dependence, are naturally defined by
imagining a perfect static triangular lattice of particles:
$$
E^{\rm cold}/N = e^{\rm cold} = (30/\pi)(1-r)^3 \ ; \
P^{\rm cold}V/N = -(dE^{\rm cold}/dV) = (45/\pi)r(1 - r)^2 \ .
$$
$$
\rho \dot e^{\rm cold} =  -\nabla u:P^{\rm cold} \ .
$$
Here $r$ is the separation of the six nearest neighbors in a cold
triangular lattice, so that $\rho = \sqrt{4/3}/r^2$.

Just as in the equilibrium Gr\"uneisen model the thermal energy and the
nonviscous part of the thermal pressure are taken to be proportional to
temperature:
$$
e^{\rm thermal} = c(K_x + K_y)/N \ ; \
P^{\rm thermal} = \gamma \rho e^{\rm thermal} \ ,
$$
where $\gamma$ is Gr\"uneisen's constant and $ck$ is a heat capacity.

The Krook-Boltzmann relaxation terms, with relaxation time $\tau$, are the
simplest means for guaranteeing thermal equilibrium, with the two temperatures
approaching one another far from the shockfront.

Because molecular dynamics simulations indicate that temperature becomes a
tensor in strong shockwaves, a tentative two-temperature formulation can be
based on separating the internal energy and the pressure into the three
components suggested by classical statistical mechanics, including
Newtonian shear and bulk viscosities:
$$
E = Ne = \Phi^{\rm cold} + \Phi^{\rm thermal} + K_x + K_y \ ;
$$
$$
P_{xx} = P_{eq} - (\eta + \eta_V)du/dx \ ; \
P_{yy} = P_{eq} + (\eta - \eta_v)du/dx \ ;
$$
$$
P_{eq} = 
\rho[\phi^{\rm cold} + \gamma ckT_{xx}] \ {\rm or} \
\rho[\phi^{\rm cold} + \gamma ckT_{yy}] \ {\rm or} \
\rho[\phi^{\rm cold} + \gamma ck(T_{xx} + T_{yy})/2] \ ;
$$
$$
e^{\rm thermal} = \phi^{\rm thermal}+ (k/2)(T_{xx} + T_{yy})
= ck(T_{xx} + T_{yy}) \ .
$$

The sum of the three energy evolution equations just given is designed to
reproduce the usual First Law energy equation,
$$
\dot E = \dot E_Q - \dot E_W \ ,
$$
where $\dot E_Q$ and $\dot E_W$ are the comoving rates at which heat enters
the fluid and at which the fluid performs work on its surroundings.  The
constitutive relations for $P$ and $Q$ must also be given.  For a
two-dimensional Newtonian fluid with shear viscosity $\eta $ and
bulk viscosity $\eta _v$ we have
$$
  P_{xx} = P_{\rm eq} - (\eta + \eta _v) du/dx \ ;
\ P_{yy} = P_{\rm eq} + (\eta - \eta _v) du/dx \ .
$$
The heat flux is given by a generalization of Fourier's law, with
independent contributions from $\nabla T_{xx}$ and $\nabla T_{yy}$.

Additional generalizations of this approach can  be developed as needed
to describe results from simulations.  It is only required that any
such model satisfy energy conservation and reduce to the 
Navier-Stokes-Fourier model in the weak-shock limit. To illustrate the
possibilities, compare the molecular dynamics results of Figure 3 to the
model calculations of Figure 9.  In Figure 9 the relaxation
time has been increased to 3, the heat capacity doubled, to $ck=2k$, and
the heat conductivity set equal to 6 so as to
better match the empirical results of molecular dynamics.  The value
of Gr\"uneisen's $\gamma $ is 0.3, and the bulk and shear viscosities
are both equal to unity.  The results from these choices (which are by no
means optimized) resemble the shockwave profiles obtained with molecular
dynamics.

\section{Conclusions and Problems for the Future}

We have shown here that it is relatively easy to model the thermal
anisotropicity found in atomistic simulations of strong shockwaves.
Thermal relaxation, bulk viscosity, and Gr\"uneisen equations of state
are useful components of a kinetic shockwave model. 
By apportioning the longitudinal and transverse thermal portions of
the work, heat, and heat flux vector a variety of useful models can
be developed and used to reproduce results from simulations.  A
forward-in-time fourth-order Runge-Kutta (as opposed to backward Euler)
integration of the cell and nodal motion equations results in accurate
and stable continuum dynamics.

One of the recent observations from molecular dynamics is that the stress
and heat flux lag somewhat behind the strainrate and the temperature
gradient\cite{b12}.  It is desirable that models be generalized to
reflect these lags.  Some study of time-delayed differential equations
is necessary to model this phenomenon.

A significant goal is the extension of these same ideas to the
fluctuating stress and heat flows of two and three dimensional fluids.
A comparison of results from molecular dynamics with those from two
and three-dimensional two-temperature continuum simulations should
provide useful tools for describing fluctuations within the overall
one-dimensional flows.

These results show that even far-from-equilibrium shocks can be treated
in a semiquantitative way by relating the tensor parts of the energy
flows to one another in a relatively simple way.  An intriguing result
of some model calculations is the stable reversal of the direction of the
heat flux vector.  Though this reversal seems unphysical, there is no
difficulty in obtaining stable numerical profiles which include flux reversal.

\begin{figure}
\includegraphics[height=9cm,width=5.25cm,angle=-90]{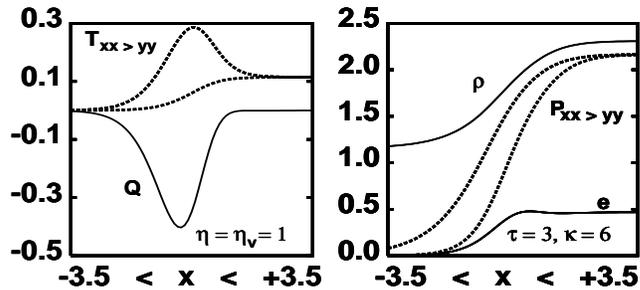}
\caption{
Solution of the {\em generalized} Navier-Stokes-Fourier equations
with both heat and work contributing solely to $T_{xx}$ and with the heat flux
$Q = -\kappa(\nabla T_{xx} + 7\nabla T_{yy})/8$. The shear viscosity,
bulk viscosity, heat conductivity, and thermal relaxation times are
respectively 1, 1, 6, and 3.  Gr\"uneisen's $\gamma $ is 0.3 and $ck=2k$.
}
\end{figure}

The thermodynamic irreversibility of the shockwave process has an interest
independent of the definition of temperature and is worth futher study. The
shock process itself obeys purely Hamiltonian mechanics, and Liouville's
Theorem\cite{b20}.  Even so, by using Levesque and Verlet's integer version
of the leapfrog algorithm\cite{b21} the entire shockwave dynamics can 
be precisely reversed, to the very last bit.  The apparent paradox,
a perfectly time-reversible but thermodynamically irreversible process,
can most clearly be illustrated by simulating the (inelastic) collision
of two zero-pressure blocks of fluid.  The collision of the blocks, with
velocities $\pm u_p$ generates two shockwaves, with velocities
$\pm(u_s-u_p)$.  Two snapshots from such a simulation are shown in Figure 10.

\begin{figure}
\includegraphics[height=18cm,width=10.5cm,angle=-90]{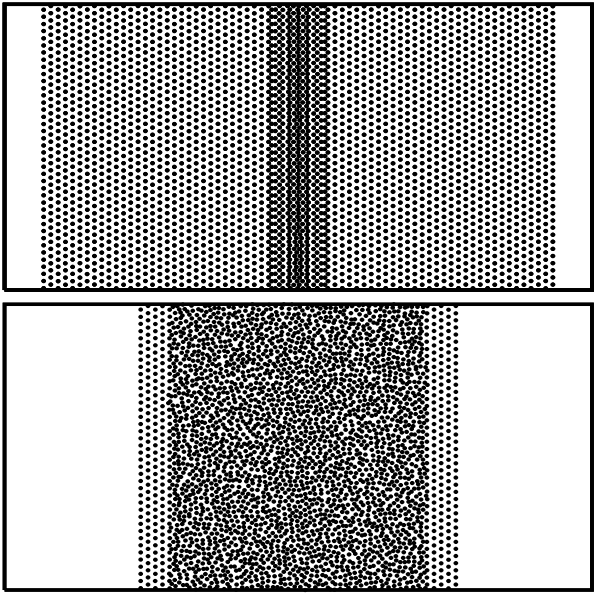}
\caption{
Two snapshots of the collision of two 1600-particle slabs (periodic in
the $y$ direction, with height 20 and initial width 20$\sqrt{3/4}$.
The initial velocities, $u_p = \pm0.965$ give twofold shock compression,
followed by a nearly isentropic free expansion at the free surfaces.
}
\end{figure}

\section{Acknowledgments}

We appreciate stimulating comments from several colleagues: Paco Uribe,
Vitaly Kuzkin, Howard Brenner, Michel Mareschal,  Krzysztof Wojciechowski,
and Jim Lutsko.  Brad Holian and Michel Mareschal have provided continuing
inspiration, through their emails and seminal publications.


\begin{thebibliography}{99}

\bibitem{b1}  H. M. Mott-Smith, ``The Solution of the Boltzmann Equation
              for a Shockwave'', Physical Review {\bf 82}, 885-892 (1951).

\bibitem{b2}  D. Gilbarg and D. Paolucci, ``The Structure of Shockwaves
              in the Continuum Theory of Fluids'', Journal of Rational
              Mechanics Analysis {\bf 2}, 617-642 (1953).

\bibitem{b3}  L. D. Landau and E. M. Lifshitz, {\em Fluid Mechanics}
              (Pergamon, Oxford, 1959).  Chapter IX is devoted to shockwaves.

\bibitem{b4}  R. E. Duff, W. H. Gust, E. B. Royce, M. Ross, A. C. Mitchell,
              R. N Keeler, and W. G. Hoover, ``Shockwave Studies in 
              Condensed Matter'', pp. 397-406 in {\em Behavior of Dense
              Media under High Dynamic Pressures} (Gordon and Breach,
              New York, 1968).

\bibitem{b5}  V. Y. Klimenko and A. N. Dremin, ``Structure of Shockwave
              Front in a Liquid'', pages 79-83 in {\em Detonatsiya,
              Chernogolovka} (Akademia Nauk, Moscow, 1978).

\bibitem{b6}  W. G. Hoover, ``Structure of a Shockwave Front in a Liquid'',
              Physical Review Letters {\bf 42}, 1531-1534 (1979).

\bibitem{b7}  B. L. Holian, W. G. Hoover, B. Moran, and G. K. Straub,
              ``Shockwave Structure {\em via} Nonequilibrium Molecular
              Dynamics and Navier-Stokes Continuum Mechanics'',
              Physical Review A {\bf 22}, 2798-2808 (1980).

\bibitem{b8}  B. L. Holian, ``Modeling Shockwave Deformation {\em via}
              Molecular Dynamics'', Physical Review A {\bf 37}, 2562-2568
              (1988).

\bibitem{b9}  O. Kum, Wm. G. Hoover, and C. G. Hoover, ``Temperature
              Maxima in Stable Two-Dimensional Shockwaves'', Physical
              Review E {\bf 56}, 462-465 (1997).

\bibitem{b10} F. J. Uribe, R. M. Velasco, and L. S. Garc\'ia-Col\'in, ``Two
              Kinetic Temperature Description for Shock Waves'', Physical
              Review E {\bf 58}, 3209-3222 (1998).

\bibitem{b11} K. Xu and E. Josyula, ``Multiple Translational Temperature
              Model and its Shock Structure Solution'', Physical Review
              E {\bf 71}, 056308 (2005).

\bibitem{b12} Wm. G. Hoover and C. G. Hoover, ``Tensor Temperature and 
              Shockwave Stability in a Strong Two-Dimensional Shockwave'',
              Physical Review E {\bf 80}, 011128 (2009).

\bibitem{b13} Wm. G. Hoover and C. G. Hoover, ``Shockwaves and Local
              Hydrodynamics; Failure of the Navier-Stokes Equations'',
              Condensed Matter arXiv:0909.2882.

\bibitem{b14} B. L. Holian and M. Mareschal, ``A New Heat-Flow Equation
              Motivated by the Ideal-Gas Shockwave'', Physical Review E
              (submitted, 2009).

\bibitem{b15} Wm. G. Hoover and C. G. Hoover, ``Nonlinear Stresses and
              Temperatures in Transient Adiabatic and Shear Flows 
              {\em via} Nonequilibrium Molecular Dynamics: Three
              Definitions of Temperature'', Physical Review E {\bf 79},
              046705 (2009).

\bibitem{b16} Wm. G. Hoover, C. G. Hoover, and J. F. Lutsko, ``Microscopic
              and Macroscopic Stress with Gravitational and Rotational
              Forces'', Physical Review E {\bf 79}, 036709 (2009).

\bibitem{b17} L. S. Garc\'ia-Col\'in and M. S. Green, ``Definition of
              Temperature in the Kinetic Theory of Dense Gases'',
              Physical Review {\bf 150}, 153-158 (1966).

\bibitem{b18} A. L. Garcia, M. M. Mansour, G. C. Lie, and E. Clementi,
              ``Numerical Integration of the Fluctuating Hydrodynamic
              Equations'', Journal of Statistical Physics {\bf 47},
              209-228 (1987).

\bibitem{b19} A. Puhl, M. M. Mansour, and M. Mareschal, ``Quantitative
              Comparaison of Molecular Dynamics with Hydrodynamics in
              Rayleigh-B\'enard systems'', Physical Review A {\bf40}
              1999-2012, (1989).

\bibitem{b20} Wm. G. Hoover, ``Liouville's Theorems, Gibbs' Entropy,
              and Multifractal Distributions for Nonequilibrium Steady
              States'', Journal of Chemical Physics {\bf 109},
              4164-4170 (1998).

\bibitem{b21} O. Kum and W. G. Hoover, ``Time-Reversible Continuum
              Mechanics'', Journal of Statistical Physics {\bf 76},
              1075-1081 (1994).

\end{thebibliography}
\end{document}